\documentclass[fleqn,12pt,twoside]{article}
\usepackage[headings]{espcrc1}

\readRCS
$Id: espcrc1.tex,v 1.2 2004/02/24 11:22:11 spepping Exp $
\usepackage{graphicx}
\usepackage[figuresright]{rotating}

\newcommand{\AmS}{{\protect\the\textfont2
  A\kern-.1667em\lower.5ex\hbox{M}\kern-.125emS}}

\title{Heavy-to-light ratios as a test
of medium-induced energy loss at RHIC and the LHC}

\author{\underline{N. Armesto}\address[USC]{Dep. F\'{\i}sica de
Part\'{\i}culas and IGFAE,
Universidade de Santiago de
Compostela,
Spain}%
        \thanks{Investigador Ram\'on y Cajal of Ministerio de Educaci\'on y
Ciencia
of Spain; financial support of CICYT of Spain under project
FPA2002-01161 is acknowledged.},
M. Cacciari\address[PVI]{LPTHE, Universit\'e Pierre et Marie Curie (Paris 6),
France},
A. Dainese\address[Padova]{Universit\`a degli Studi di Padova and INFN,
Padova, Italy},
C.A. Salgado\address[CERN]{Department of Physics, CERN, Theory Division,
CH-1211 Gen\`eve 23, Switzerland} and U.A. Wiedemann\addressmark[CERN]}
       
\runtitle{Heavy-to-light ratios as a test
of medium-induced energy loss at RHIC and the LHC}
\runauthor{N. Armesto, M. Cacciari, A. Dainese, C.A. Salgado and U.A.
Wiedemann}

\begin{document}

\maketitle

\begin{abstract}
The ratio of nuclear modification factors of high-$p_T$ heavy-flavored mesons to
light-flavored hadrons (heavy-to-light ratio) is shown to be a sensitive
tool to test medium-induced energy loss at RHIC and LHC energies.
Heavy-to-light ratios of $D$ mesons at RHIC in the region $7<p_T<12$ GeV, and
of $D$ and $B$ mesons at the LHC in the region $10<p_T<20$ GeV, are
proposed for such a test. Finally, the different contributions to
the nuclear modification factor for electrons at RHIC are
analyzed. Preliminary PHENIX and STAR data are compatible with
radiative energy loss provided the contribution of electrons from beauty
decays is small compared to that from charm.
\end{abstract}

\section{INTRODUCTION}
\label{intro}

Energy loss by medium-induced gluon radiation is the standard explanation
for the suppression of high transverse momentum hadron spectra
in nucleus-nucleus collisions compared to p-p collisions. The description of 
this effect (see \cite{carlinhos} and references therein)
considers the rescattering of the emitted gluon with a medium characterized by
its length $L$, and by the so-called BDMPS transport coefficient $\hat{q}$.
This transport coefficient measures the color field strength in the medium
by characterizing the amount of squared transverse momentum transferred from
the medium to the hard parton per unit path length. It is one general expectation
of parton energy loss that due to their larger color charge, gluons will
radiate more than light quarks which, due to their vanishing mass, will radiate
more than heavy quarks~\cite{Dokshitzer:2001zm,Djordjevic:2003zk}.

Heavy-to-light ratios $R_{D(B)/h}=R^{D(B)}_{AA}/R^h_{AA}$ are mainly 
sensitive to two effects. First, in hadronic collisions at sufficiently high center
of mass energy, light hadrons are predominantly produced by gluons, while
heavy mesons have quark parents. $R_{D(B)/h}$ increases due to this
color charge effect. Second, heavy quarks radiate less energy than light ones
- this mass effect increases $R_{D(B)/h}$ further. There are additional
smaller effects which affect  $R_{D(B)/h}$ and which are discussed in \cite{Armesto:2005iq}.

Our model \cite{Armesto:2005iq} is based on a LO  pQCD formalism, as implemented in
the PYTHIA event generator \cite{Sjostrand:2001yu}, supplemented by radiative energy
loss via quenching weights~\footnote{Publicly available FORTRAN routines to
compute quenching weights in the massive case can be found in
www.pd.infn.it/$^\sim$dainesea/qwmassive.html.}, which give the probability
that a parton traversing a medium characterized by $\hat{q}$ and $L$ loses a given
fraction of its energy. The geometry of the medium is modeled realistically
\cite{Dainese:2004te}. After their quenching, partons are fragmented
according to standard procedures, see \cite{Armesto:2005iq}.

\section{HEAVY-TO-LIGHT RATIOS}
\label{htlr}

\subsection{RHIC}
\label{htlrrhic}

In order to get predictions at RHIC energies, the value of the only medium-dependent
model parameter, $\hat{q}$, is fixed from the high-$p_T$ suppression of light hadrons 
in central Au--Au at $\sqrt{s_{NN}}=200$ GeV. Values of $\hat{q}$ lie
\cite{Dainese:2004te} within the range $4\div 14$ GeV$^2$/fm. In
\cite{Armesto:2005iq}, the description of experimental data in d--Au collisions
and the predictions for the nuclear ratios of $D$-mesons and electrons from
charm at RHIC, are presented.

\begin{figure}[htb]
\vskip -0.5cm
\begin{center}
\includegraphics[width=9cm,height=7.cm]{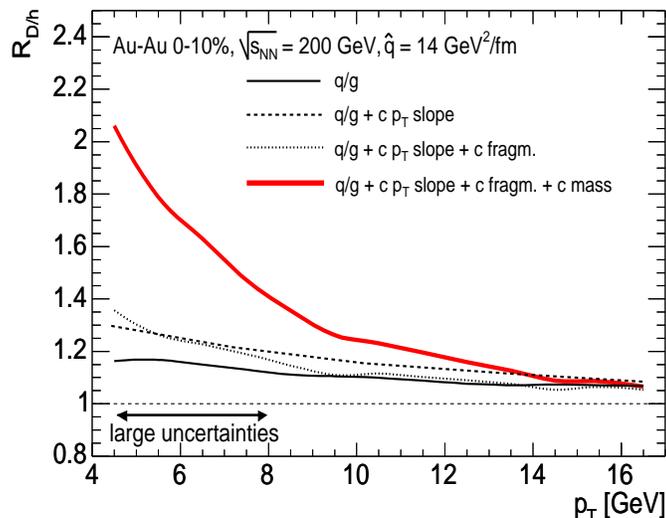}
\end{center}
\vskip -1.7cm
\caption{Different contributions to the heavy-to-light ratio of $D$ mesons
in central (0--10\%) Au--Au collisions at $\sqrt{s_{NN}}=200$ GeV. Different
curves correspond to the case in which the charm distribution is described
i) with the same $p_T$ spectrum, fragmentation function and parton energy loss
as a light quark, ii) with a realistic charm $p_T$ spectrum only, iii) with
the
charm $p_T$ spectrum and fragmentation function of a realistic charm quark
and iv) for the realistic case including the mass dependence of parton
energy loss.}
\label{figrhic}
\end{figure}

\vskip -0.6cm
Here we show in Fig. \ref{figrhic} the
heavy-to-light ratio of $D$ mesons, with the different ingredients discussed
in the previous Section selectively switched on. The mass effect turns out to
be the largest
in the region $p_T<7\div 8$ GeV, a region subject to large
uncertainties due to the possibility of fragmentation in medium. For light
hadrons this is indicated by the baryon-to-meson anomaly, and heavy quarks
could be more affected due to their slower movement through the medium. On the
other hand, in the region $7<p_T<12$ GeV both the color charge and the mass
effects play a similar quantitative role.

\subsection{LHC}
\label{htlrlhc}

To extrapolate the value of $\hat{q}$ for different energies,
a proportionality of the density with the multiplicity is assumed.
Extrapolations from RHIC to LHC lie in the range $2.5\div 7$
\cite{Dainese:2004te,Armesto:2004ud}. Conservatively, a wide range
$\hat{q}=4$, 25 and 100 GeV$^2$/fm is explored \cite{Armesto:2005iq}.

\begin{figure}[htb]
\vskip -0.6cm
\begin{center}
\includegraphics[width=10cm,height=8cm]{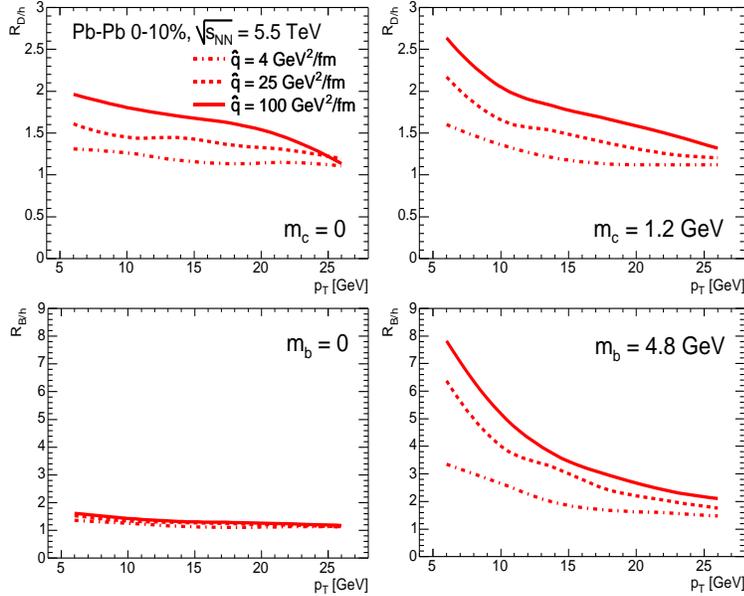}
\vskip -1.7cm
\end{center}
\caption{Heavy-to-light ratios in central Pb--Pb at the LHC
for $D$  (upper plots) and $B$ mesons
(lower
plots) for a realistic heavy quark mass (plots on the right) and
for a case
in which the quark mass dependence of parton energy loss is neglected
(plots on the left).}
\label{figlhc}
\end{figure}

\vskip -0.8cm
Results for nuclear ratios
of $B$ and $D$ mesons and decay electrons can be found in
\cite{Armesto:2005iq}. Here in Fig. \ref{figlhc} the heavy-to-light ratios of
$B$  and $D$ mesons are shown, with explicit indication of
the effect of the mass on the quenching. $R_{D/h}$ is sensitive mainly to
the color charge effect, which is more pronounced than at RHIC due to the
smaller Bjorken $x$. $R_{B/h}$ is very
sensitive to the mass effect. Both can be seen in the window
$10<p_T<20$ GeV, which should be safe from the influence of a medium on
fragmentation.

\section{NUCLEAR MODIFICATION FACTOR FOR ELECTRONS AT RHIC}
\label{electrons}

\begin{figure}[htb]
\vskip -0.5cm
\begin{center}
\includegraphics[height=6.2cm,width=7.75cm]{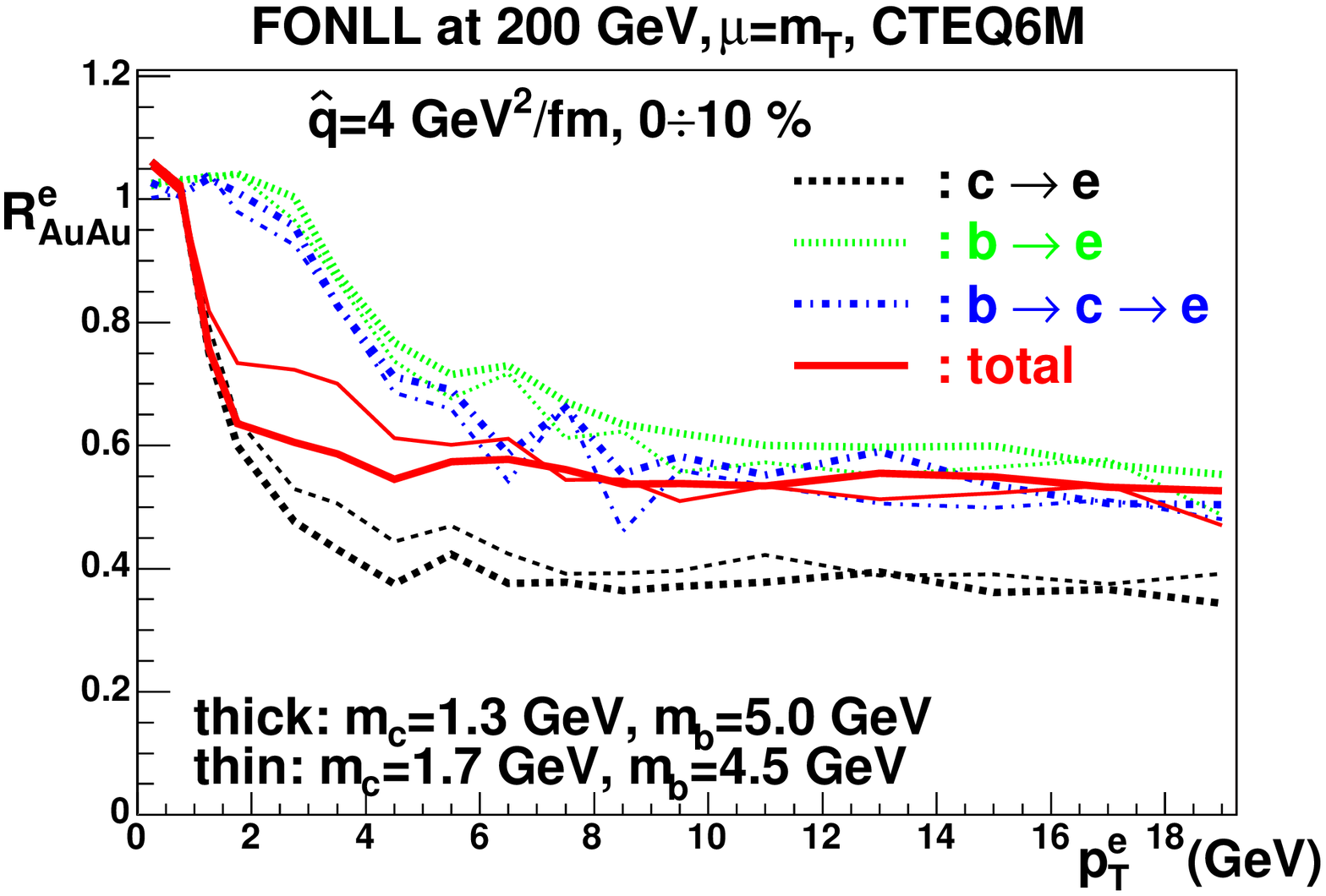}\hfill\includegraphics[height=6.2cm,width=7.75cm]{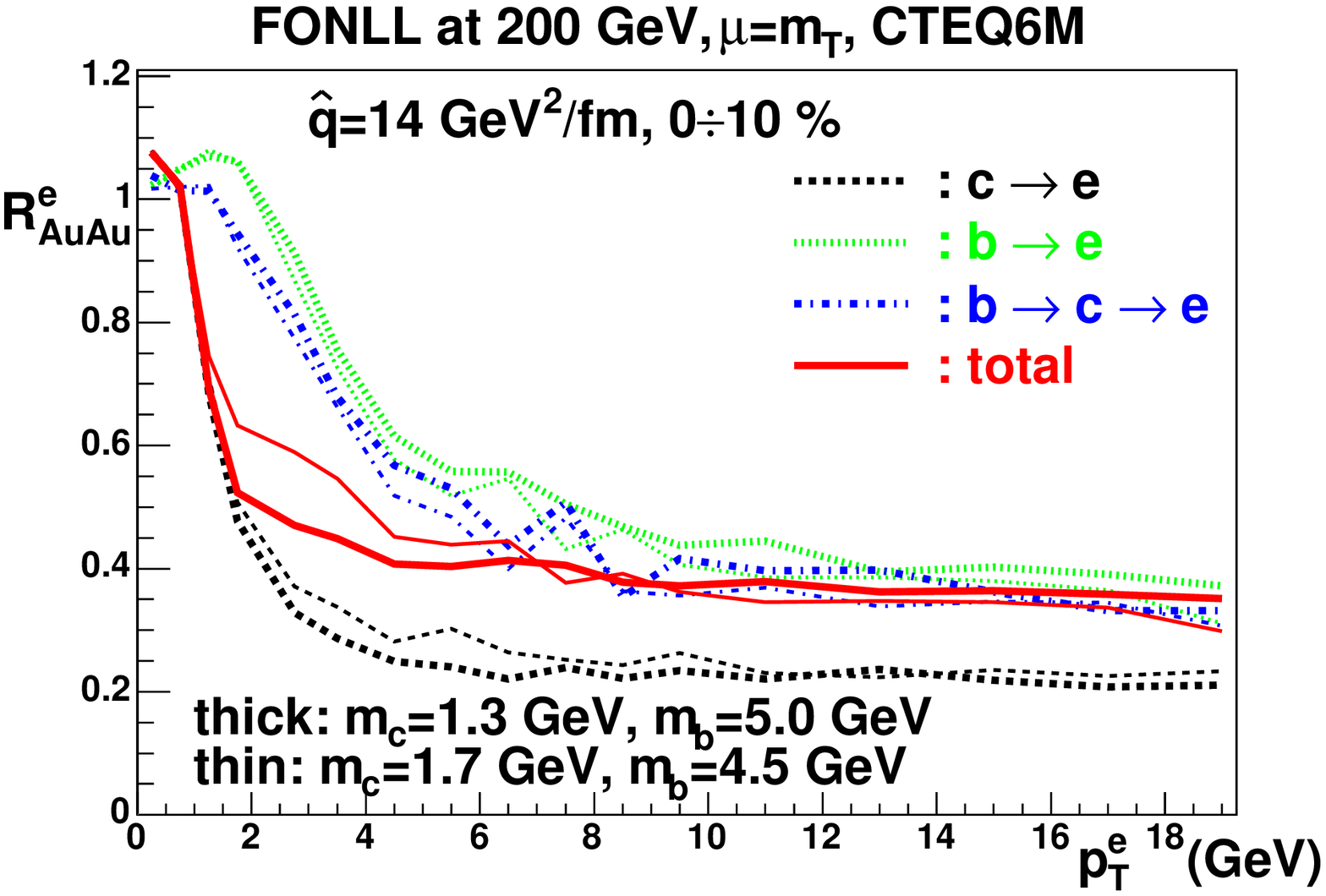}
\end{center}
\vskip -1.5cm
\caption{Contributions at RHIC to $R_{AA}^e$ for electrons coming
from heavy meson decay for $\hat{q}=4$ (plot on the left) and 14 (plot on the
right) GeV$^2$/fm, for different quark masses.}
\label{figrate}
\end{figure}

With heavy-flavor
mesons not directly reconstructed in nucleus-nucleus collisions at
RHIC, the present discussion is focused on nuclear ratios for
non-photonic electrons $R_{AA}^e$
\cite{Adcox:2002cg}. Preliminary measurements by PHENIX and STAR
\cite{phenixstar} show a large suppression by a factor $3\div 5$,
$R_{AuAu}^e\sim 0.2\div 0.4$,
in the region $4<p_T^e<10$ GeV, which has been considered as
incompatible with radiative energy loss \cite{Djordjevic:2005db} and/or
indicative of heavy quark thermalization \cite{eiqui}.
As discussed in \cite{Djordjevic:2005db}, bottom quarks
should be less suppressed than charm quarks due to their higher mass,
and their contribution to the electron spectra should tend to increase the
corresponding nuclear ratio.
In order to analyze this point, we have supplemented state-of-the-art
pQCD
computations of heavy flavor production \cite{Cacciari:1998it} (FONLL: fixed
order NLO plus NLL resummation) with energy loss via quenching weights
\cite{Armesto:2005iq}. Full details and a discussion of
uncertainties will
be presented elsewhere \cite{noso}. Here we show in Fig. \ref{figrate}
the different heavy flavor
contributions to $R_{AA}^e$,
with the uncertainty in FONLL
given by a variation in heavy quark masses.
These results are compatible with preliminary data \cite{phenixstar} provided
the contribution of electrons from beauty decays with respect to charm is
smaller than predicted by FONLL.
Thus an
estimation
of the uncertainties \cite{noso} in charm \cite{phenix2} and
beauty spectra at RHIC energies becomes crucial. 
                                                                                
\vskip -0.6cm
\noindent{\bf Acknowledgments:}
We thank the organizers for such a nice and fruitful conference.


\begin{thebibliography}{99}
\bibitem{carlinhos}
C.~A.~Salgado,
arXiv:hep-ph/0510062;
X.N.~Wang, these proceedings.
\bibitem{Dokshitzer:2001zm}
Y.L.~Dokshitzer and D.E.~Kharzeev,
Phys.\ Lett.\ B 519  (2001) 199.

\bibitem{Djordjevic:2003zk}
M.~Djordjevic and M.~Gyulassy,
Nucl.\ Phys.\ A 733 (2004) 265;
B.W.~Zhang, E.~Wang and X.N.~Wang,
Phys.\ Rev.\ Lett.\  93 (2004) 072301;
N.~Armesto, C.A.~Salgado and U.A.~Wiedemann,
Phys. Rev. D 69 (2004) 114003.

\bibitem{Armesto:2005iq}
N.~Armesto, A.~Dainese, C.A.~Salgado and U.A.~Wiedemann,
Phys.\ Rev.\ D 71 (2005) 054027.

\bibitem{Sjostrand:2001yu}
T. Sjostrand, L. Lonnblad and S. Mrenna,
arXiv:hep-ph/0108264.
%
\bibitem{Dainese:2004te}
A.~Dainese, C.~Loizides and G.~Paic,
Eur.\ Phys.\ J.\ C 38 (2005) 461;
K.J.~Eskola, H.~Honkanen, C.A.~Salgado and U.A.~Wiedemann,
Nucl.\ Phys.\ A 747 (2005) 511.

\bibitem{Armesto:2004ud}
N.~Armesto, C.A.~Salgado and U.A.~Wiedemann,
Phys.\ Rev.\ Lett.\  94 (2005) 022002.

\bibitem{Adcox:2002cg}
PHENIX Coll.: K.~Adcox {\it et al.}, 
Phys.\ Rev.\ Lett.\  88 (2002) 192303;
S.S.~Adler {\it et al.},
Phys.\ Rev.\ Lett.\  94 (2005) 082301.

\bibitem{phenixstar} PHENIX Coll.:
S.~A.~Butsyk,
arXiv:nucl-ex/0510010;
STAR
Coll.:
J.~Bielcik, these proceedings.; H.-B.~Zhang, {\it ibid}.

\bibitem{Djordjevic:2005db}
M.~Djordjevic, M.~Gyulassy, R.~Vogt and S.~Wicks,
arXiv:nucl-th/0507019; M.~Djordjevic, these proceedings.

\bibitem{eiqui} R.F.~Rapp, these proceedings; D.A.~Teaney, {\it ibid}.;
  B.~Zhang, L.~W.~Chen and C.~M.~Ko,
  arXiv:nucl-th/0509095.

\bibitem{Cacciari:1998it}
M.~Cacciari, M.~Greco and P.~Nason,
JHEP 9805 (1998) 007;
M.~Cacciari, P.~Nason and R.~Vogt,
Phys.\ Rev.\ Lett.\ 95 (2005) 122001; R.~Vogt, these proceedings.

\bibitem{noso}
N.~Armesto, M.~Cacciari, A.~Dainese, C.A.~Salgado and U.A.~Wiedemann, in
preparation.

\bibitem{phenix2} PHENIX Coll.: Y.~Kwon,
  arXiv:nucl-ex/0510011.

\end{thebibliography}
\end{document}